\documentclass[prl, twocolumn, groupedaddress]{revtex4-1}
\usepackage{amsmath}
\usepackage{amssymb}
\usepackage{graphicx}
\usepackage{tikz}

\newcommand{\n}{{\bf n}}
\newcommand{\ncn}{{\bf n} \cdot \nabla \times {\bf n}}

\begin{document}
\title{Contact Topology and the Structure and Dynamics of Cholesterics}
\author{Thomas Machon}
\affiliation{Department of Physics and Astronomy, University of Pennsylvania, 209 South 33rd Street, Philadelphia, Pennsylvania 19104, USA}
\begin{abstract}
Using tools and concepts from contact topology we show that non-vanishing twist implies conservation of the layer structure in cholesteric liquid crystals. This leads to a number of additional topological invariants for cholesteric textures, such as layer numbers, that are not captured by traditional descriptions, characterises the nature and size of the chiral energy barriers between metastable configurations, and gives a geometric characterisation of cholesteric dynamics in any context, including active systems, those in confined geometries or under the influence of an external field.
\end{abstract}

\maketitle

\section{Introduction}

Cholesteric liquid crystals often display a high number of metastable configurations~\cite{bouligand74,Dierking}, particularly in confined geometries, such as droplets~\cite{bouligand84,zhou16,sec12, xu97}, shells~\cite{darmon16b, darmon16, tran17}, colloidal systems~\cite{tkalec11,jampani11,machon13}, or systems confined between parallel plates~\cite{ackerman14,chen13,ackerman17,smalyukh10} where the interplay between chirality, elasticity and surface interactions can produce a large variety of structures. These configurations typically contain recurring structural motifs, such as folded and contorted layers, pitch defects, dislocations and double-twist cylinders (vortex tubes) that may be found in, for example, blue phases~\cite{wright89} and cholesteric fingers~\cite{baudry98}. A characterisation of defects in these structures can be given in terms of pitch-axis descriptions of the cholesteric phase~\cite{volovik77, priest74} which can be formalised by defining defects in the pitch axis as singularities or degeneracies, such as umbilic lines, in the gradient tensor of the director field~\cite{beller14,machon16}. These defects, which are found at the centre of double-twist cylinders or where the layer structure of a cholesteric changes, can be locally characterised in this way. The theory, however, does not extend rigorously to give a defect algebra describing full defect combination rules or establishing topological equivalence between configurations. The pitch axis depends on the gradients of the director field, they cannot be varied independently and so the standard defect algebra based on the homotopy groups of $SO(3)/D_{2d}$ cannot be applied~\cite{beller14,mermin79}. 

Beyond defects, there are many other apparently topological properties of a cholesteric texture that are not captured locally by degeneracies or singularities. As a simple example, for a helical texture of a cholesteric between two parallel plates, the `layer number', the number of $\pi$ twists in the director groundstate \eqref{eq:gs} appears to be conserved and changes only through the presence of defects, as in the case of Grandjean-Cano wedge geometries~\cite{grandjean21,cano68,smalyukh02}. Despite this, the layer number in a cholesteric is not defined in terms of the director field by any standard topological theory in condensed matter.

Here we propose the language of contact topology as the natural one in which to discuss topological aspects of three-dimensional chiral structures such as cholesteric liquid crystals. Using tools and concepts from contact topology, we show how the robustness of these structures, such as layers and double-twist cylinders, as well as the high degree of metastability in cholesterics follows from the tendency of the cholesteric to twist, expressed as an energetic preference for the unit magnitude director field $\n$ to satisfy $\ncn \neq 0$. In particular, we show how this implies that the dynamics of the cholesteric conserve the layer structure of the texture, appropriately defined. We also show how additional topological invariants, such as layer numbers and a property of `overtwistedness' that gives a topological stability criterion for  solitons with zero Hopf charge~\cite{chen13,ackerman17,bouligand78}, can be defined in the regime of non-vanishing twist. The theory we describe gives homotopy invariants of cholesteric textures in this regime originating in contact topology~\cite{geiges}, allowing topological classification of cholesteric configurations in a way that cannot be accomplished in standard theories. We show that changes in the layer structure of the cholesteric can only be achieved by passing through a configuration with vanishing twist, allowing us to characterise the chiral energy barriers that ensure the stability of these systems. Under this condition, we derive the geometric evolution equation that describes the motion of cholesteric layers and double-twist cylinders. This equation holds in any regime, including systems in confined geometries, active systems~\cite{whitfield17} and those with applied fields, as long as the geometric condition $\ncn \neq 0$ is satisfied throughout. While contact topology has enjoyed some application to the hydrodynamics of Beltrami fields and Euler flows~\cite{etnyreghrist_series} (a connection we also highlight) as well as to wave propagation~\cite{Arnold}, to our knowledge there has not, as yet, been an application of the theory to condensed matter systems.

\section{The Conservation of Layer Structure in Cholesterics}

In the one-elastic-constant approximation, the bulk Frank free energy for a cholesteric in a domain $\Omega$ can be written in terms of the unit-magnitude director field $\n$ as
\begin{equation}
F = \frac{K}{2}\int_\Omega (\nabla \cdot \n)^2 + (q_0 + \n \cdot \nabla \times \n)^2 + ((\n \cdot \nabla) \n)^2 dV,
\label{eq:frank}
\end{equation}
from which one observes that sufficient conditions for a texture to be a global energy minimiser are zero splay ($\nabla \cdot \n =0$) and bend ($(\n \cdot \nabla ) \n =0$), but non-zero twist, with $\ncn = -q_0$. These are met by the helical texture
\begin{equation}
\n_{gs} = \big ( \cos q_0 z, \sin q_0 z, 0 \big ),
\label{eq:gs}
\end{equation}
which is unique up to proper Euclidean transformations. The twist, $\ncn$, has geometric significance through the Fr\"{o}benius integrability condition, and is zero if and only if $\n$ is locally the normal vector to a family of surfaces. From a geometric perspective cholesterics are materials for which $\n$ has an energetic preference for non-integrability (observe that a smectic liquid crystal can be characterised in precisely the opposite sense). Such non-integrable configurations are chiral, the chirality determined by the sign of twist, with $\ncn<0$ for a right-handed texture. In this paper, we will impose the contact condition, $\ncn \neq 0$, on the cholesteric and study the topological classes of texture that arise from this restriction, as well as the consequences for possible dynamics. The contact condition is local and, in spirit at least, the topological information obtained by imposing this restriction can be thought of as similar to obtaining topological invariants in condensed matter systems by demanding that a system is gapped throughout momentum space.

For simplicity, we will restrict ourselves to the study of textures in simply connected domains with $\n$ non-singular, so that $\n$ may be considered as a genuine unit magnitude vector field. In terms of the 1-form $n$ dual to $\n$, the contact condition may be written as $n \wedge dn \neq 0$, which in flat space is numerically equal to $\ncn$. 1-forms satisfying this condition (or more properly the tangent planes orthogonal to them) such as $n$ constitute examples of contact structures~\cite{geiges}, maximally non-integrable hyperplane fields. It is then natural to investigate the degree to which the tools and concepts of contact topology may be brought to bear on the study of cholesterics. We note that in the following it is important, for purely technical reasons which we explain below, to distinguish the 1-form $n$ from the director field $\n$.

Given the identification of cholesterics and contact structures, the naive approach is to consider all possible director fields satisfying $\ncn \neq 0$ and study their equivalence classes up to isotopy, so that two director fields are considered topologically equivalent if one may be reached from another without breaking the contact condition. This problem has yielded many deep results mathematically, perhaps principal among them is Eliashberg's classification of overtwisted contact structures~\cite{eliashberg89} (see below for a definition) which states that the isotopy classes of certain contact structures are equivalent to the homotopy classes of maps into $\mathbb{R} \mathbb{P}^2$, or nematic textures from a liquid crystalline perspective. There are other results, such as Yutaka's classification~\cite{yutaka97} which states that up to isotopy, helical (in the sense of \eqref{eq:gs}) contact structures in a box with periodic boundary conditions are specified up to isotopy by a layer number and a Miller plane determining the layer normals (mathematically equivalent to a primitive element of $H_2(T^3;\mathbb{Z})$). A more esoteric result with potential applications to cholesterics is Honda's classification~\cite{honda00,honda00b} of contact structures on $T^2 \times [0,1]$, which applies directly to the study of cholesterics in toroidal shells with planar anchoring, where he obtains an elaborate classification involving continued fraction expansions of twisting numbers on the boundary.

While classification results such as these are useful (we will discuss elementary applications below) it turns out that contact topology has things to say even about the local structure of cholesterics and their time evolution, as well as the evolution of geometric structures within the cholesteric, such as layers. The analogy between cholesterics and the layered system of smectics is oft-made; indeed their layered ground states share the same elasticity theory~\cite{deGennes}. However, in the case of the smectic, the layer structure is clearly defined through the smectic mass-density wave whereas for the cholesteric the layer structure is difficult to define unambiguously for a general configuration~\cite{beller14}. Despite this ambiguity, we will show below that for a very general definition of cholesteric layers, non-zero twist implies that the layer structure in a cholesteric is conserved. The immediate corollary is that for the layer structure to change, one must pass through a configuration containing a region of zero twist. In the smectic case such a conformational change is necessarily accompanied by the creation of defects, where the smectic order parameter vanishes. In the cholesteric case an entirely analogous role is played by twist, which acts as a local cholesteric order parameter. 

\begin{figure*}
\begin{center}
\begin{tikzpicture}
    \node[anchor=south west,inner sep=0] at (0,0) {\includegraphics{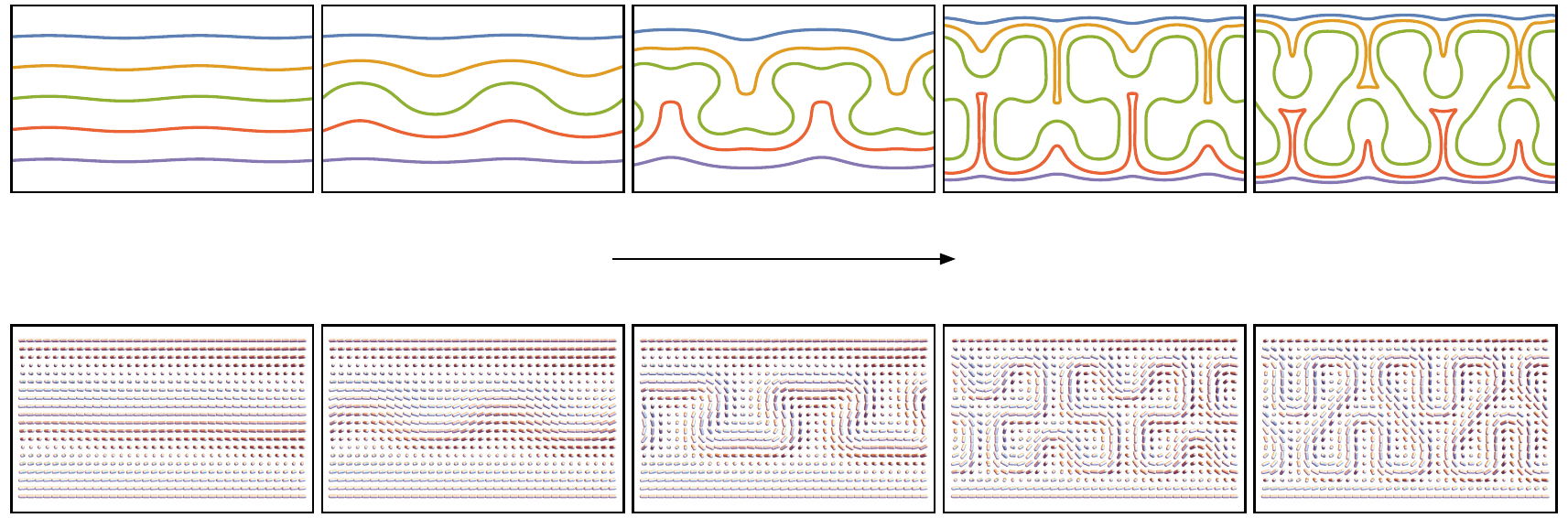}};
    \node[anchor=center] at (8.7,3.25) 
    {Layers flow along {\bf w}};
     \node[anchor=center] at (8.7,2.5) 
    {Director evolution};
\end{tikzpicture}
\end{center}
\caption{{\it Left to Right}: Simulated evolution of the Helfrich-Hurault instability in a thin cholesteric sheet under gradient descent dynamics. Observe that the equilibirum configuration (far right) is not the groundstate of the system, which consists of three $2 \pi$ twists. There is a chiral barrier of size $K_2 q_0^2/2$ that cannot be overcome by gradient descent. {\it Top}: Evolution of layers, defined as surfaces with normal $\nu$ satisfying $\nu \cdot {\bf n}=0$ in the initial state. The layers evolve according to the equation $\partial_t \Sigma(u,v) = {\bf w}(\Sigma(u,v),t)$, this preserves the layer condition, and so describes the evolution of the layer structure of the cholesterics and is well defined as long as the twist doesn't vanish.  {\it Bottom}: Evolution of the director field according to gradient descent.} 
\label{fig:movie}
\end{figure*}

To establish this we consider a foundational result in contact topology: Gray's theorem~\cite{gray59}. As applied to cholesterics, it states that if a system described by $n_0$ evolves to another configuration $n_1$, with $n_t \wedge dn_t \neq 0$, $t\in[0,1]$, then there is a coordinate transformation (diffeomorphism) relating $n_0$ and $n_1$. Defining time-indexed coordinates ${\bf x}_t$, this implies we can find a relationship ${\bf x}_t({\bf x}_0)$ such that
\begin{equation}
n_i({\bf x}_t, t) d x_t^i = \alpha_t n_i({\bf x}_0, 0) d x_0^i.
\label{eq:ct}
\end{equation}
Here we find the reason for the aforementioned distinction between $n$ and $\n$. \eqref{eq:ct} says that the director field at time $t$ can be related to the director field at time 0 by a coordinate transformation, if we assume $n$ transforms as a 1-form. One can write the equivalent equation in terms of the vector field $\n$, but one would require the inclusion of metric factors associated to the coordinate transformation. This statement makes no requirement of the physical transformation properties of the director field, it is a geometric construction that allows us to explore cholesteric topology. To proceed we differentiate \eqref{eq:ct}, finding
\begin{equation}
\partial_t n_i + w^j \partial_j n_i + n_j \partial_i w^j = \lambda_t n_i,
\label{eq:lie}
\end{equation}
where ${\bf w} = d {\bf x}_t/dt$ and $\lambda_t = d \log \alpha_t / dt$. Establishing the claim at the beginning of the paragraph then amounts to showing that one may always find a vector field ${\bf w}$ solving \eqref{eq:lie} as long as $\ncn \neq 0$~\cite{note_integrating}. Assuming $w^in_i=0$ allows \eqref{eq:lie} to be made into an algebraic equation for $\bf w$ and taking the cross-product with $\n$ on both sides yields the solution
\begin{equation}
 {\bf w}   = \frac{1}{\big( \ncn \big )} \n \times \partial_t \n,
 \label{eq:w}
\end{equation}
which tightly constrains the structure and dynamics of cholesterics. As long as $\ncn \neq 0$, the geometric flow field ${\bf w}$ is well-defined and governs the evolution of structures in the texture such as layers and double-twist cylinders, that evolve in time by flowing along ${\bf w}$. While ${\bf w} $ can be defined for any cholesteric, in general it will not correspond to the physical flow field. As an aside it is amusing to note that \eqref{eq:lie}, which contains the Lie derivative of $n$, can be obtained as a limit of the Ericksen-Leslie (EL) equations. A small amount of algebraic manipulation allows the EL equations for the director field to be written as
\begin{equation}
\partial_t n^i +c_1\mathcal{L}_{\bf v} n^i + c_2\delta^{ij} \mathcal{L}_{\bf v} n_j -\frac{1}{\gamma_1} h^i = \lambda n^j,
\label{eq:erick}
\end{equation}
where $h^i$ is the molecular field, $c_1+c_2=1$ and $c_1-c_2 = -\gamma_2 / \gamma_1$, with $\gamma_1$ and $\gamma_2$ the nematic viscosities.  The Lie derivatives $\mathcal{L}_{\bf v} n^i = v^j \partial_j n^i - n^j \partial_j v^i$ and  $\mathcal{L}_{\bf v} n_i = v^j \partial_j n_i + n_j \partial_i v^j$ describe how the the nematic flows under the influence of the velocity field ${\bf v}$. The relative size of $c_1$ and $c_2$ controls the degree to which ${\bf n}$ flows as a vector $(c_1=1)$ or a 1-form $(c_2=1)$. By setting the viscosity ratio $\gamma_1 / \gamma_2=1$ then taking the limit $\gamma_1 \to \infty$, \eqref{eq:erick} becomes equivalent to \eqref{eq:lie}, so that in this limit the cholesteric flow field itself generates a diffeomorphism representing the evolution of ${\bf n}$.

The flow field ${\bf w}$ describes the geometric evolution of the cholesteric. If the twist does not vanish, so that ${\bf w}$ is well defined, then this implies that the geometric structures in the cholesteric must be preserved by time evolution; in other words: non-vanishing twist implies conservation of the layer structure in cholesteric liquid crystals. To see this in more detail we will give a definition of a cholesteric layer. While it is clear that in the helical groundstate, the `layers' are normal to the pitch axis, in more general contexts defining layers in this way cannot be done. A particularly informative example is the twist-bend nematic texture, given by $\n_{hc} = \cos \theta \n_{gs} + \sin \theta {\bf e}_z$, where $\n_{gs}$ is the helical phase, \eqref{eq:gs}. This texture has $\ncn\neq 0$, and naively one would still want to define the `layers' as level sets of $z$, however such a texture does not have a pitch axis pointing in the vertical direction~\cite{beller14}, so that the definition of the layers cannot be made in this way.

To evade this issue we will use a very general definition of a cholesteric layer. An embedded measuring surface $\Sigma$ in the material will be considered a layer if $\n$ is never normal to $\Sigma$, so that $\n$ projected into the tangent space of $\Sigma$ is always non-zero. It is easy to see that the level sets of $z$ then correspond to layers in both the helical phase and the heliconal phase. We note that this definition of a layer is (intentionally) very general, and that a small perturbation to a layer will typically still correspond to a layer so that one does not have a unique foliation, even locally, of the system by layers. 

Given a layer $\Sigma$, one may project $\n$ into the tangent space of $\Sigma$, defining a so-called `characteristic foliation'~\cite{geiges}, $\mathcal{F}$, of $\Sigma$ whose leaves are the integral curves of $\n$ restricted to $\Sigma$. The layer condition then amounts to $\mathcal{F}$ having no singular points. There is a second natural foliation $\overline{\mathcal{F}}$ defined as the directions in the tangent planes to $\Sigma$ that are orthogonal to $\n$. These observations amount to the fact that at each point on $\Sigma$, one may find two directions, ${\bf e}_1$ and ${\bf e}_2$ such that 
\begin{equation}
 {\bf e}_1 \cdot \n >0,\quad {\bf e}_2 \cdot \n =0.
\label{eq:conditions}
\end{equation}
If $\Sigma$ is allowed to flow along ${\bf w}$ under time evolution, then it is simple to show that \begin{equation}
D_t({\bf e}_i \cdot \n) = \lambda_i {\bf e}_i \cdot \n,
\end{equation} where $D_t = \partial_t + {\bf w} \cdot \nabla$, so that the conditions \eqref{eq:conditions} are preserved and $\Sigma$ is still a layer by our definition. It follows, therefore, that the topological layer structure of a cholesteric is preserved as long as twist does not vanish. 

\begin{figure}
\begin{center}
\begin{tikzpicture}
    \node[anchor=south west,inner sep=0] at (0,0) {\includegraphics{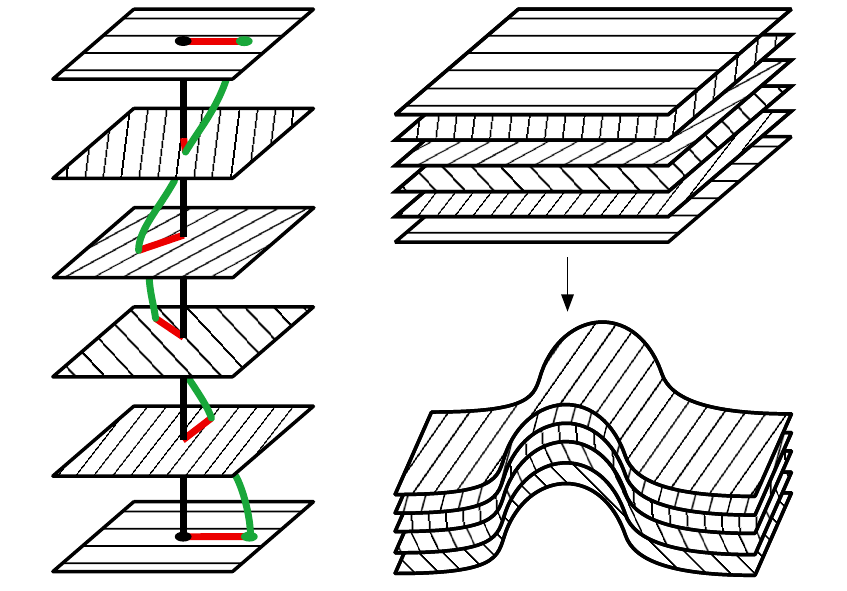}};
     \node[anchor=center] at (6.15,3.12) 
    {${\bf w}$};
      \node[anchor=center] at (0.2,5.6) 
    {(a)};
       \node[anchor=center] at (4,5.6) 
    {(b)};
\end{tikzpicture}
\end{center}
\caption{(a): Defining the Thurston-Bennequin number. The Legendrian loop $\gamma$ (black) is pushed along $\n$ (red) to form $\gamma^\prime$ (green), the Thurston-Bennequin number is defined as $\textrm{Lk}(\gamma, \gamma^\prime)$. (b) If the twist does not vanish, then the existence of layers is preserved, their deformation is controlled by the vector field ${\bf w}$, along which they flow.} 
\label{fig:link}
\end{figure}
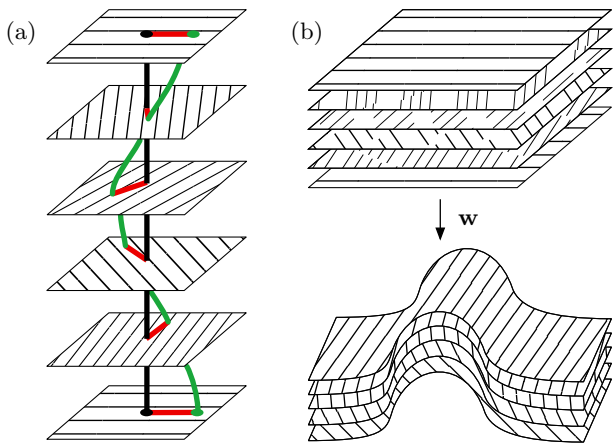

An application of \eqref{eq:w} can be seen in the Helfrich-Hurault instability, a layer buckling instability of the helical phase (Fig.~\ref{fig:movie}) under an applied electric field~\cite{helfrich71,hurault73} or mechanical strain~\cite{clark73}. Here we present a simple quasi-two-dimensional simulation, similar to that considered in Ref.~\cite{watson00}, to illustrate these concepts. We consider a cholesteric between two vertically separated parallel plates, with fixed boundary conditions, $\n = {\bf e}_x$ on the top and bottom~\cite{note_sim}. The system is initialised on a 150x300 grid in a sinusoidally perturbed frustrated helical phase with one $2\pi$ twist with $q_0$ chosen such that the equilibrium number of $2 \pi$ twists is three. Time evolution is given by gradient descent of \eqref{eq:frank} which has been found to be a good approximation to experimentally observed dynamics in this context~\cite{watson00}. We have
\begin{equation}
\partial_t \n = -\frac{1}{\gamma_1}P_{\n} \cdot \frac{\delta F}{\delta {\bf n}},
\label{eq:gradient_descent}
\end{equation}
where $(P_{\n})_{ij} = \delta_{ij} - n_i n_j$ is the projection operator orthogonal to $\n$. As the simulation proceeds, the system exhibits the buckling instability, the frustrated twist energy relaxes through the growth of the perturbation, forming a contorted set of layers, the evolution of which is prescribed by ${\bf w}$. The layer structure is preserved as long as the twist does not vanish. Contact topology allows us to estimate the chiral barrier associated to a layer reconnection process. At a layer reconnection the local free energy density $f$ would necessarily satisfy
\begin{equation}
f \geq K_2 q_0^2/2.
\label{eq:chiral}
\end{equation}
If we assume that such a layer reconnection process happens locally in some region $R$ of volume $V$, with $\n$ constant outside $R$, then gradient descent dynamics implies that the free energy within $R$ is decreasing over time, so that there must be sufficient splay and bend energy to overcome the chiral barrier. In particular, if $s =\int_R (\nabla \cdot \n)^2$ and $b =\int_R ((\n \cdot \nabla )\n)^2$, then we must have $K_1 s + K_3 b \gtrapprox K_2 q_0^2 V$, since vanishing twist bounds the free energy from below by $K_2q_0^2$. The inequality is approximate since we do not assume twist vanishes throughout the entirety of $V$.  It follows from this that the layer structure may only change in regions of sufficiently high bend or splay. In this example, we see the origins of the high number of metastable configurations of cholesterics -- any change in the texture that alters the layer topology must pass through a chiral energy barrier. 

\section{Layer Number Invariants}

There is another property of the example in Fig.~\ref{fig:movie} that can be explained using contact topology. Note that even as the system reaches an equilibrium (local free energy minimum) it does not find the groundstate, corresponding to a helical texture with three $2 \pi$ twists, instead settling to a periodic configuration of contorted layers. In larger systems~\cite{watson99,watson00} the metastable state may evolve further to reach a configuration where large regions of the system have reached the equilibrium configuration, but there nevertheless remain soliton-like folded layer structures and pitch defects, stabilised by the chiral energy density barrier of $K_2 q_0^2/2$ required to change the layer structure. the aggregate collection of these structures is not topological in a nematic (their aggregate charge is zero), where all helical textures with an even number of $\pi$ twists are in the same topological class~\cite{note_gareth}, nor are they topological in a biaxial nematic (or pitch axis description of a cholesteric), where the same statement holds. This is true simply because the configuration originated in a helical phase, with no defects, so that by running time backward, one can make the configuration defect-free. Because of this, these textures cannot be topologically distinguished using homotopy theory, however from the perspective of contact topology they are distinct. Imposing the condition $\ncn \neq 0$ leads to additional topological invariants of textures -- allowing, for example, one to distinguish configurations with differing numbers of layers on topological grounds. These invariants are defined through the properties of measuring loops and surfaces. If $\gamma$ is an embedded arc in the domain with $\gamma^\prime \cdot {\bf n} =0$ (a so-called Legendrian arc), then allowing $\gamma$ to flow with ${\bf w}$ ensures that $\gamma$ remains an embedded Legendrian arc under time evolution. Indeed, the normalised tangent vector $\gamma^\prime$ satisfies the vorticity equation (with a Lagrange multiplier):
\begin{equation}
D_t \gamma^\prime = (\gamma^\prime\cdot \nabla) {\bf w } + \lambda \gamma^\prime,
\label{eq:vort}
\end{equation}
so that $ D_t \big (\n \cdot \gamma^\prime \big ) = 0$ if $\gamma^\prime\cdot \n =0$ initially, the orthogonality condition $\n \cdot \gamma^\prime=0$ is then preserved under flow along ${\bf w}$. This allows one to derive topological invariants of closed Legendrian loops and for our purposes the most relevant is the Thurston-Bennequin number $tb(\gamma)$~\cite{geiges} (Fig.~\ref{fig:link}(a)). Given a Legendrian loop $\gamma$, one can push $\gamma$ an amount $\epsilon$ along ${\bf n}$ to form a new arc $\tilde \gamma$. We then define $tb(\gamma) = \textrm{Lk}(\gamma, \tilde{\gamma})$ which is a topological invariant of $\gamma$, preserved under time evolution. We note that in this regard the topology of cholesterics has a similar flavour to the conservation of helicity and linking of vortex lines in Euler flows~\cite{moffatt69,arnold74}. 

The Thurston-Bennequin number allows one to distinguish helical phases with different numbers of layers. Consider a cholesteric between two parallel plates of separation $h_0$, so that the domain is $\Omega = \mathbb{R}^2 \times [0,h_0]$, with boundary conditions ${\bf n}|_{\partial \Omega} = \pm(1,0,0)$. In this system, the possible helical textures are specified by an integer, $l$: \begin{equation}\n_l = \left (\cos \frac{\pi l z}{h_0},  \sin \frac{\pi l z}{h_0},  0 \right),\label{eq:helical}\end{equation}
with the minimum energy helical configuration given by the integer that minimises $(\pi l - q_0 h_0)^2$. 

We now show that there is no family of director fields connecting $\n_k$ and $\n_l$  ($l\neq k$), satisfying the boundary conditions, that does not pass through a configuration with a region of zero twist. Using \eqref{eq:frank}, the free energy density of such an untwisted region can be bounded from below as $f \geq K q_0^2/2$, and as we saw above (Fig.~\ref{fig:movie}), this barrier is sufficient to effectively forbid the transition. 

To obtain the result, one considers the set of all unknotted Legendrian curves $\gamma$ going between the two plates. For any such $\gamma$, one can define a half-integer modified Thurston-Bennequin number $tb(\gamma)$ which gives the number of times $\n$ winds around $\gamma$. For a vertical path $\gamma$, $2 tb(\gamma) = l$, the layer number and for any other unknotted path $\tilde \gamma$, the twisting number satisfies the inequality~\cite{note1,yutaka97} 
\begin{equation}
tb(\tilde \gamma) \geq l/2.
\label{eq:twist}
\end{equation}
Establishing this result is technical, and requires embedding a neighbourhood of $\gamma$ into the tight contact structure on $S^3$. It follows from \eqref{eq:twist} that the minimum $tb$ number over all allowed (unknotted) paths is an invariant of the texture and equal to $l/2$, half the layer number. In particular, this invariant is sufficient to topologically distinguish the initial $2\pi$ twist configuration in Fig.~\ref{fig:movie} from the equilibrium $6 \pi$ twist configuration. There was a sleight of hand in this argument. Mathematically it is possible, without violating the contact condition, for the flow of ${\bf w}$ to send points to infinity in finite time, so that the measuring line $\gamma$ is broken at infinity. Physically, however, no such transition can be achieved without infinite energetic cost or distortions propagating from the boundary in a finite system. An example of such a deformation would be to create a non-singular $\lambda^+ \lambda^-$ dislocation at infinity and drag it through the system, increasing the layer number by 2. To achieve such a deformation in finite time would require $\partial_t \n$ to diverge, costing infinite kinetic energy. Mathematically such deformations may be forbidden by demanding that ${\bf w}$ integrates to a diffeomorphism of the domain, a so-called proper isotopy. 

Finally, we note for completeness that it should also be possible to construct $tb(\gamma)$ and establish a similar result using the natural Frenet-Serret apparatus that one may associate to the pitch axis~\cite{beller14}, so that $tb(\gamma)$ becomes a generalisation of a Frenet-Serret self-linking number, preserved as long as $Tr(\chi)\neq 0$, where $\chi_{ij} = \epsilon_{ljk}n_k \partial_i n_l$ is the chirality tensor~\cite{beller14,efrati14,machon16}. 

\begin{figure}
\begin{center}
\includegraphics{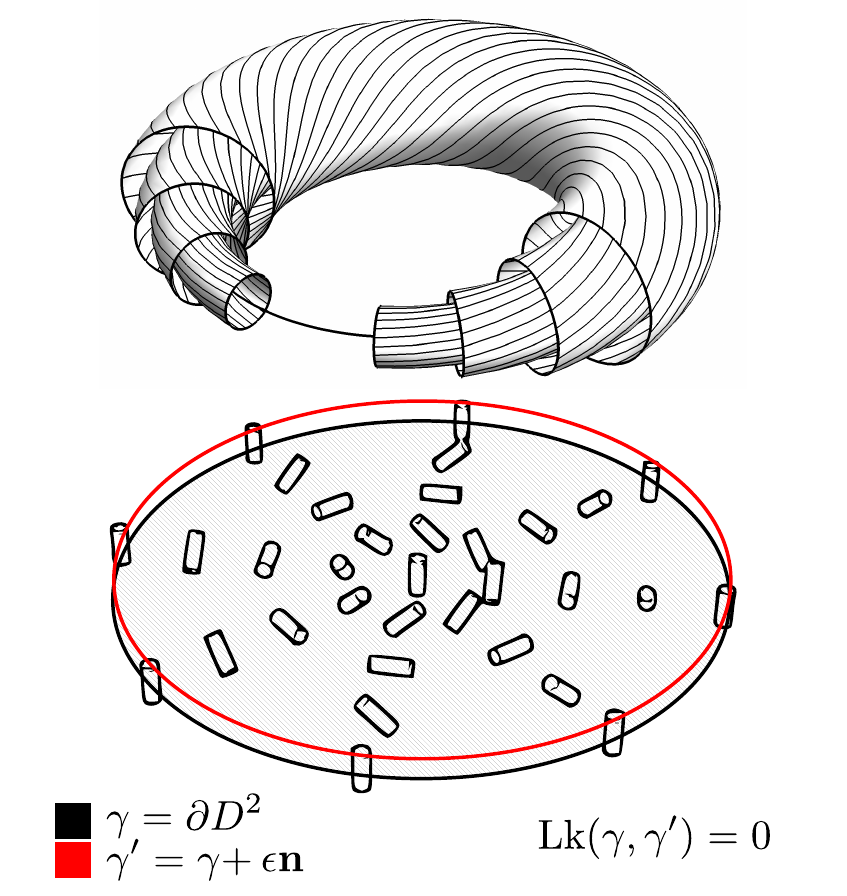}
\end{center}
\caption{{\it Top}: A double-twist cylinder, consisting of cylindrical layers. This layer structure is preserved under time evolution, with the layers flowing along ${\bf w}$. If the director rotates a full $\pi$ to the boundary, forming an overtwisted disk, then the existence of the double-twist cylinder is a topological invariant of the texture. {\it Bottom}: An overtwisted disk, $D$, defined as an embedded disk whose boundary  $\gamma$ is Legendrian and has $tb(\gamma )=0$. The existence of such a disk is a topological invariant of a cholesteric.} 
\label{fig:dtc}
\end{figure}

\section{Double Twist Cylinders and Overtwisted Disks}

So far, we have only focused on layers, but entirely analogous results hold for double-twist cylinders, where the layers are cylindrical or toroidal (by the Poincar\'{e}-Hopf theorem the layer condition \eqref{eq:conditions} can only be satisfied on a closed surface if it is a torus). In the same way that additional topological invariants can be defined for layered structures, they can also be defined for double-twist cylinders of a sufficient radius.

Of particular interest from the perspective of contact topology are double-twist cylinders that contain a full $\pi$ rotation of the director field, examples of which can be seen as vortices in Skyrmion lattices. Mathematically one can define such cylinders topologically. If $\gamma$ is a closed Legendrian loop bounding a disk, $D$, with $tb(\gamma)=0$ then $\gamma$ is said to be the boundary of a cross-sectional disk of a double-twist cylinder with a full $\pi$ rotation of $\n$, or an overtwisted disk in the mathematical literature. An example is illustrated in Fig.~\ref{fig:dtc}. Observe that an overtwisted disk implies the existence of an entire tube. Pushing the disk $D$ along its normal creates a new disk $D^\prime$, but it is easily verified by considering the boundary that $D^\prime$ is also an overtwisted disk.

In the contact topology literature, contact structures with such vortex tubes are termed overtwisted, distinguished from tight contact structures, such as \eqref{eq:gs}, which do not possess overtwisted disks. This dichotomy is deep, isotopy classes of overtwisted contact structures are equivalent to homotopy classes of director fields~\cite{eliashberg89} whereas isotopy classification of tight contact structures is often complicated. Despite this, the presence, or otherwise, of an overtwisted disk (and the corresponding vortex tube) is a topological invariant of the texture -- if such a disk can be found in a texture, then the texture is a topological soliton in a cholesteric.

\begin{figure}
\begin{center}
\includegraphics{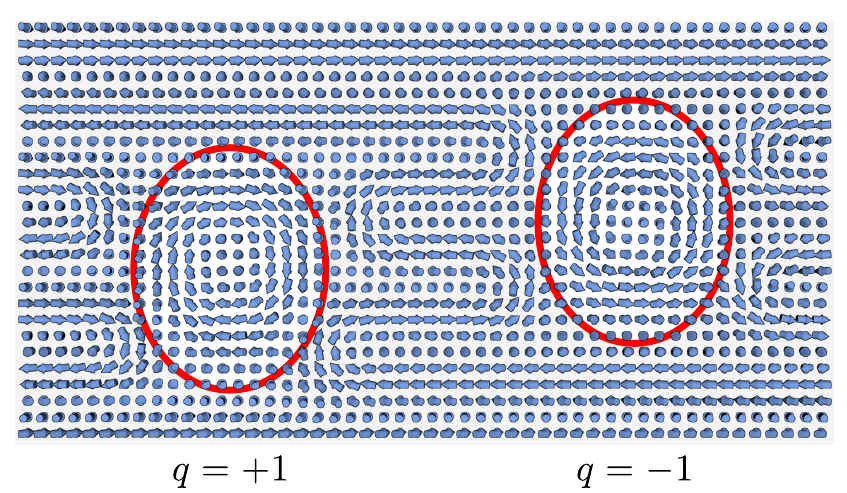}
\end{center}
\caption{Two overtwisted disks (red) in a helical texture. Each disk encloses a Skyrmion of indicated charge. One may consider this configuration as two Skyrmion tubes embedded in a helical texture extending out of the page. While the two have net charge zero, the configuration is topologically protected by the condition of non-zero twist as it possesses overtwisted disks (note that due to the discrete nature of the simulated texture and plot, exact locations for the overtwisted disks cannot be shown).} 
\label{fig:skyrmion}
\end{figure}

The traditional picture of an overtwisted disk (Fig.~\ref{fig:dtc}) is identical to the well-known twist Skyrmion distortion. Skyrmions are topological in a nematic, representing charge $1$ in the non-trivial homotopy group $\pi_2(\mathbb{R}\mathbb{P}^2)=\mathbb{Z}$, so one might wonder what additional information one gains from the contact topology perspective in this case. The Skyrmion charge changes sign under $\n \to - \n$ (the origin of much of the topological subtlety in nematics) but chirality, as measured by $\ncn$, is invariant (hence the allowed presence of the chiral term in the Frank free energy). As such, cholesterics can possess twist Skyrmions of either charge. One can then consider a configuration with Skyrmions of total charge $0$, as indicated in Fig.~\ref{fig:skyrmion}. Such a configuration is not topological in terms of Skyrmion charge. However, contact topology supplies us with the additional overtwistedness invariant, which identifies the configuration as topologically protected.

This distinction between tight and overtwisted has further application in the study of non-singular topological solitons in three dimensional cholesteric liquid crystals. Typically these are characterised by the Hopf invariant $\nu$~\cite{chen13,bouligand78,ackerman17}, an integer originating in the homotopy group $\nu \in\pi_3(\mathbb{R}\mathbb{P}^2)$, which can be measured by computing the linking of pre-images (the Pontryagin-Thom construction~\cite{chen13}). Configurations with $\nu=0$ are therefore not topological when measured this way. However, in a cholesteric we obtain an additional topological invariant of overtwistedness that can distinguish these configurations, resulting in the existence of two topologically distinct classes of configuration with $\nu =0$. The standard helical texture does not contain an overtwisted disk, and so textures with overtwisted disks are topologically distinct from those without. It follows that cholesteric textures containing overtwisted disks but with $\nu=0$ form a new class of topological solitons in cholesterics (note that if $\nu \neq 0$ then it is known that all configurations must contain an overtwisted disk). These statements follow from Eliashberg's classification~\cite{eliashberg93} of contact structures on $\mathbb{R}^3$.

We have shown that cholesterics admit a new class of topological soliton -- configurations that are overtwisted (containing double-twist cylinders) but have Hopf invariant zero, $\nu=0$. A reasonable question to ask is what such a soliton would look like? While we do not present any experimental data here, it is likely that they have been observed. The toron configuration~\cite{smalyukh10} as well as more recent work on Hopf solitons~\cite{ackerman17} show many stable soliton configurations with $\nu=0$, containing an overtwisted disk in the vertical projection. While care must be taken in these cases as the frustrated boundary conditions have zero twist, it is likely that, by using a relative perspective, for example taking advantage of Eliashberg's results for the classification of contact structures in a ball with fixed boundary~\cite{eliashberg93}, the configurations may be shown to represent the overtwisted topological class, and so obtain topological status in the cholesteric. Furthermore, we speculate that the Hopf-soliton anti-Hopf-soliton pair observed numerically in a simulation of a ferromagnet with Dzyaloshinskii-Moriya (DM) coupling~\cite{borisov11} also corresponds to such a soliton. While chiral ferromagnets are a different system to cholesterics, the theory presented here directly applies to ferromagnetic systems with DM coupling also, assuming smooth evolution of the magnetisation. More abstractly, one can construct a topological example of overtwisted configurations with arbitrary Hopf charge using a cut-and-paste procedure beginning with the helical phase, and then performing a standard operation, known as a Lutz twist~\cite{geiges}, that inserts double-twist cylinders along appropriately chosen transverse curves, $\gamma$, that satisfy $\gamma^\prime \cdot {\bf n}\neq 0$. The Hopf charge, $\nu$, of the resulting configuration is equal to the total linking number (including the contact topological self-linking number) of the transverse curve~\cite{geiges}, allowing one to specify an arbitrary value of $\nu$.

\section{Conclusion}

In this paper we have only sketched the connection between contact topology and cholesterics, and there is a great deal of future work that may be done on this relationship. For example, the theory of bypasses in contact topology~\cite{honda00,honda00b} should be able to be adapted to give a theory of dislocations in cholesterics, and the study of non-coorientable contact structures should give insight into some global properties of singular disclination lines. There is also the prospect of deriving more sophisticated invariants for cholesteric textures based on contact homology theories, which would likely have their application to the topology of knotted or linked defect lines in cholesterics. We must also stress that there is, obviously, much that the contact topology description cannot capture. The standard homotopy theory of defects assumes only a continuous map into the groundstate manifold to derive its results. The theory of contact topology makes just one more assumption, namely that this map is differentiable, so that the 3-form $n \wedge d n$ may be defined. In this sense, the theory is `one step up' from the basic homotopy theory. Of course, because of this, it does not depend on the metric of the domain, in particular the theory is insensitive to geometric properties such as length-scales, curvatures and so on, in the same way that the homotopy theory is insensitive to these properties. 

\begin{acknowledgments}
I would like to thank G.P.\ Alexander, R.D.\ Kamien and L.\ Tran for helpful comments and for reading an early version of this manuscript. It is a pleasure to acknowledge additional conversations with R.B.\ Kusner, J.E.\ Etnyre and J.\ Nelson. This work was supported by NSF Grant DMR-1262047, the Simons Foundation as well the EPSRC through grant No.\ A.MACX.0002 and the University of Warwick through a Chancellor's International Scholarship and an IAS Early Career Fellowship.
\end{acknowledgments}


\begin{thebibliography}{10}
\bibitem{Dierking} I. Dierking {\it Textures of Liquid Crystals}, (Wiley-VCH; Weinheim, 2003).
\bibitem{bouligand74} Y. Bouligand, J. Phys. (Paris) {\bf 35}, 959 (1974).
\bibitem{zhou16}  Y. Zhou, E. Bukusoglu, J.A. Martinez-Gonzalez, M. Rahimi, T.F. Roberts, R. Zhang, X. Wang, N.L. Abbott and J.J. de Pablo, ACS Nano {\bf 10}, 6484 (2016).
\bibitem{bouligand84} Y. Bouligand and F. Livolant, J. Physique {\bf 45}, 1899 (1984).
\bibitem{sec12} D. Se\v{c}, T. Porenta, M. Ravnik and S. \v{Z}umer, Soft Matter {\bf 8}, 11982 (2012).
\bibitem{xu97} F. Xu and P. P. Crooker, Phys. Rev. E {\bf 56}, 6853 (1997).

\bibitem{darmon16b} A. Darmon, M. Benzaque, D. Se\v{c}, S. \v{C}opar, O. Dauchota and T. Lopez-Leon, Proc. Natl. Acad. Sci. USA {\bf 113}, 9469 (2016).
\bibitem{darmon16} A. Darmon, M. Benzaquen, S. \v{C}opar, O. Dauchota and T. Lopez-Leon, Soft Matter {\bf 12}, 9280 (2016).
\bibitem{tran17} L. Tran, M.O. Lavrentovich, G. Durey, A. Darmon, M.F. Haase, N. Li, D. Lee, K.J. Stebe, R.D. Kamien and T. Lopez-Leon, {\it Under Review} (2017).


\bibitem{tkalec11} U. Tkalec, M. Ravnik, S. \v{C}opar, S. \v{Z}umer and I. Mu\v{s}evi\v{c}, Science {\bf 333}, 62 (2011).
\bibitem{jampani11} V.S.R. Jampani, M. \v{S}karabot, M. Ravnik, S. \v{C}opar, S. \v{Z}umer and I. Mu\v{s}evi\v{c}, Phys. Rev. E {\bf 84}, 031703 (2011).
\bibitem{machon13} T. Machon and G.P. Alexander, Proc. Natl. Acad. Sci. USA {\bf 110}, 14174 (2013).

\bibitem{ackerman14} P.J. Ackerman, R.P. Trivedi, B. Senyuk, J. Van De Lagemaat and I.I. Smalyukh, Phys. Rev. E {\bf 90}, 012505 (2014).
\bibitem{ackerman17} P.J. Ackerman and I.I. Smalyukh, Phys. Rev. X {\bf 7}, 011006 (2017).
\bibitem{chen13} B.G. Chen, P.J. Ackerman, G.P. Alexander, R.D. Kamien and I.I. Smalyukh, Phys. Rev. Lett. {\bf 110}, 237801 (2013).
\bibitem{smalyukh10} I.I. Smalyukh,	Y. Lansac,	N.A. Clark and R.P. Trivedi, Nat. Mater. {\bf 9}, 139 (2010).
\bibitem{wright89}D.C. Wright and N.D. Mermin, Rev. Mod. Phys. {\bf 61}, 385 (1989).
\bibitem{baudry98} J. Baudry, S. Pirkl and P. Oswald, Phys. Rev. E {\bf 57}, 3038 (1998).

\bibitem{volovik77} G.E. Volovik and V.P. Mineev, Zh. Eksper. Teor. Fiz. {\bf 72},
2256 (1977); [Sov. Phys. JETP {\bf 45}, 1186 (1975).]
\bibitem{priest74} R.G. Priest and T.C. Lubensky, Phys. Rev. A {\bf 9}, 893 (1974).

\bibitem{machon16} T. Machon and G.P. Alexander, Phys. Rev. X {\bf 6}, 011033 (2016).
\bibitem{beller14} D.A. Beller, T. Machon, S. \v{C}opar, D.M. Sussman, G.P. Alexander, R.D. Kamien and R.A. Mosna, Phys. Rev. X {\bf 4}, 031050 (2014).

\bibitem{mermin79} N.D. Mermin, Rev. Mod. Phys. {\bf 51}, 591 (1979).

\bibitem{grandjean21} F. Grandjean, C. R. Acad. Sci. Paris, {\bf 172}, 71 (1921).
\bibitem{cano68} R. Cano, Bull. Soc. Fr. Min\'{e}r. Cristallogr. {\bf 91}, 20 (1968).
\bibitem{smalyukh02} I.I. Smalyukh and O.D. Lavrentovich, Phys. Rev. E {\bf 66}, 051703 (2002).


\bibitem{bouligand78} Y. Bouligand , B. Derrida, V. Po\'{e}naru, Y. Pomeau and G. Toulouse, J. Phys. France {\bf 39}, 863 (1978).
\bibitem{geiges} H. Geiges, {\it An Introduction to Contact Topology}, (Cambridge University Press; Cambridge, U.K., 2008).


\bibitem{whitfield17} C.A. Whitfield T.C. Adhyapak, A. Tiribocchi, G.P. Alexander, D. Marenduzzo and S. Ramaswamy, Eur. Phys. J. E {\bf 40}, 50 (2017).

\bibitem{etnyreghrist_series} J. Etnyre and R. Ghrist, Nonlinearity {\bf 13}, 441 (2000); Ergodic Theory Dyn. Syst. {\bf 22}, 819 (2002); Trans. Amer. Math. Soc. {\bf 352}, 5781 (2000). 
\bibitem{Arnold} V.I. Arnold, {\it Contact Geometry and Wave Propagation}, (L'Enseignement math\'{e}matique, Universit\'{e} de Gen\`{e}ve; Gen\`{e}ve, 1989); {\it Singularities of Caustics and Wavefronts} (Kluwer Academic Publishers; Dordrecht, 1990).



\bibitem{eliashberg89} Y. Eliashberg, Invent. Math. {\bf 98}, 623 (1989).
\bibitem{yutaka97} K. Yutaka, Comm. Anal. Geom. {\bf 5}, 413 (1997).
\bibitem{honda00} K. Honda, Geom. Topol. {\bf 4}, 309 (2000).
\bibitem{honda00b}K. Honda, J. Differential Geom. {\bf 55}, 83 (2000).

\bibitem{deGennes} P.G. de Gennes, {\it The Physics of Liquid Crystals}, 2nd ed.\ (Clarendon Press; Oxford, United Kingdom; 1995).

\bibitem{gray59} J.W. Gray, Ann. Math. {\bf 69}, 421 (1959).


\bibitem{note_integrating} One must also ensure that ${\bf w}$ does not have any finite-time blow-ups, so that it generates a diffeomorphism of the material domain.

\bibitem{helfrich71} W. Helfrich, J. Chem. Phys. {\bf 55}, 839 (1971). 
\bibitem{hurault73} J.P. Hurault, J. Chem. Phys. {\bf 59}, 2068 (1973).
\bibitem{clark73} N.A. Clark and R.B. Meyer, Appl. Phys. Lett. {\bf 22}, 493 (1973).

\bibitem{watson00} P. Watson, J.E. Anderson and P.J. Bos, Phys. Rev. E {\bf 62}, 3719 (2000).


\bibitem{note_sim} In one-elastic-constant approximation gradient descent the values of $\gamma_1$ and $K$ only set the timescale of the dynamics. Both values were set to $1$ for the simulation, which was run with a small enough time-step to ensure numerical stability. 


\bibitem{watson99} P. Watson, J.E. Anderson, V. Sergan and P.J. Bos, Liq. Cryst. {\bf 26}, 1307 (1999).

\bibitem{note_gareth} The fastidious reader will notice that, even in a nematic, one can topologically distinguish helical textures with an even number of $\pi$ twists from those with an odd number. This $\mathbb{Z}_2$ invariant originates in the homotopy group $\pi_1(\mathbb{R P}^2) \cong \mathbb{Z}_2$.


\bibitem{moffatt69} H.K. Moffatt, J. Fluid Mech. {\bf 35}, 117 (1969).
\bibitem{arnold74} V.I. Arnold, Sel. Math. Sov. {\bf 5} 327 (1986).


\bibitem{note1} We assume $\ncn<0$, so that the contact structure is right-handed. For a left handed contact structure one must reverse the sign of the inequality. The result can be derived by mapping $\n$ in a neighbourhood of $\gamma$ onto a standard tight contact structure in $S^3$ and observing that \eqref{eq:twist} being violated implies the contradiction that the tight contact structure contains an overtwisted disk.

\bibitem{efrati14}  E. Efrati and W.T.M. Irvine, Phys. Rev. X {\bf 4}, 011003 (2014).

\bibitem{eliashberg93} Y. Eliashberg, Internat. Math. Res. Notices {\bf 3}, 87 (1993). 


\bibitem{borisov11} A.B. Borisov and F.N. Rybakov, Low Temp. Phys. {\bf 36}, 766 (2010). 

\end{thebibliography}
\end{document}